\documentclass[12pt,a4paper]{article}
\usepackage[utf8]{inputenc}
\usepackage[english]{babel}
\usepackage{amsmath}
\usepackage{amsfonts}
\usepackage{amssymb}
\usepackage{graphicx}
\usepackage{kpfonts}
\usepackage{natbib}
\usepackage[left=2cm,right=2cm,top=2cm,bottom=2cm]{geometry}
\usepackage{xcolor}

\def\gev{\,{\rm GeV}}

\title{A Bayesian Model of\\
 Credence in Low Energy Supersymmetry}

\author{Richard Dawid and James D. Wells}
\date{}

\begin{document}
\maketitle

\begin{abstract}
We carry out a quantitative Bayesian analysis of the evolution of credences in low energy supersymmetry (SUSY) in light of the most relevant empirical data. The analysis is based on the assumption that observers apply principles of optimism or pessimism about theory building in a coherent way. On this basis, we provide a rough assessment of the current range of plausible credences in low energy SUSY and determine in which way LHC data changes those credences. For observers who had been optimistic about low energy SUSY before the LHC, the method reports that LHC data does lead to decreased credences in accordance with intuition. The decrease is moderate, however, and keeps posteriors at very substantial levels. 
{The analysis further establishes that a very high but not yet indefensible degree of pessimism regarding the success chances of theory building still results in quite significant credences in GUT and low energy SUSY for the time right before the start of the LHC. The pessimist's credence in low energy SUSY remains nearly unchanged once LHC data is taken into account.}
 
\end{abstract}

\section{Introduction}

Supersymmetry, the hypothesis that nature at a fundamental level shows a symmetry between fermionic and bosonic fields, was proposed in the early 1970s and, since then, has played a central role in high energy physics. 

 The fact that the particle spectrum observed in HEP collider experiments was not at all supersymmetric could be accounted for by assuming that SUSY is spontaneously broken at energy scales above the reach of experiments at the time. Belief in the idea that the SUSY breaking scale lies only moderately above the electroweak scale (henceforth to be called low energy SUSY) has a complicated history. 

While Supersymmetry was initially developed for conceptual reasons that were unrelated to immediate empirical considerations and thus did not suggest a specific SUSY breaking scale, it was soon understood that a number of empirical characteristics of HEP and cosmology could find a plausible explanation based on low energy SUSY. Low energy SUSY could provide a promising dark matter candidate, it made the gauge couplings meet with remarkable accuracy in line with the idea of grand unification, and a low SUSY breaking scale allowed for a stable and natural separation of scales between the electroweak scale and the Planck scale. These attractive features in conjunction with the general conceptual appeal of the SUSY hypothesis and the fact that supersymmetry was implied by string theory, the leading candidate for a theory of quantum gravity, convinced a considerable share of high energy physicists from the late 1980s onwards that low energy SUSY was likely to be found by collider experiments strong enough to test energy scales a couple of orders of magnitude above the electroweak scale.

When no SUSY signatures were found at LEP2, hopes largely remained high that it would be found at LHC. More recently, the failure to find SUSY at the LHC has led to a widespread substantial decline of credence in the hypothesis that SUSY is part of physics in the wider proximity of the electroweak scale at all. For some observers, the disappointment of high hopes to find SUSY at the LHC has led to a certain degree of skepticism regarding the usefulness of conceptual reasoning on the prospects of a theory's success in the absence of empirical confirmation. 

This paper uses tools of Bayesian reasoning to get a better handle on two questions that arise in the given situation. Is it possible to understand the rational basis for trust in low energy SUSY before the LHC? To the extent that is the case, do we, on that basis, find in the Bayesian analysis a substantial decrease of credence in SUSY based on the data collected in recent HEP experimentation?

Credence in SUSY obviously relies on a range of individual assessments and attitudes that enter a Bayesian analysis in the form of priors and likelihoods. A first step in our analysis will be to argue that those priors and likelihoods, while covering a wide range of plausible options, are less arbitrary than what one might expect at first glance. They are constrained by expectations regarding the success chances of physical theorizing that most physicists would not want to reject, as well as by the requirement of some degree of coherence of an agent's set of credences. Two general guidelines will be followed when spelling out those constraints: the experience of predictive success in the research field should be reflected in some way by sets of priors chosen for new theories; and substantial differences between priors should not be assumed without reason. Of course, these constraints still leave a wide spectrum of credences that look perfectly defensible. We will argue, however, that they turn the assessment of credences in a hypothesis like SUSY into something substantially more significant than an arbitrary number game based on boundless flexibility of subjective priors. 

The upshot of our analysis will be the following. Natural expectations regarding the scientific process lead to very substantial credence in SUSY before the failure to find it at LEP2 and the LHC. Naturalness was by no means the strongest argument for SUSY, even if one had a very strong belief in the argument's soundness. LHC data does lead to reduced credences in SUSY for a wide range of agents. But accounts that refrain from tuning credences in an implausible way only lead to moderately decreased credences that remain very substantial. A more detailed discussion of results will be provided in the conclusion.

\section{Supersymmetry at low and middle energies}

\subsection{SUSY phenomenology}

Supersymmetry first developed out of formal theoretical reasoning. In order to incorporate fermions on the two-dimensional worldsheet of string theory it was necessary to introduce supersymmetry. Supersymmetry generators were also discovered in the different context of expanding the four-dimensional spacetime symmetry algebra beyond the Poincar\'e group without violating the Coleman-Mandula theory. These formal developments percolated into the world of phenomenology, where theorists wondered if there would be observable consequences of nature possessing supersymmetry (\cite{Dreiner:2023yus}).

There were many efforts at constructing supersymmetric theories amenable to experimental pursuit. The fruit of all of these efforts came known to be called the Minimal Supersymmetric Standard Model (MSSM). The gauge symmetry of the MSSM is the same gauge symmetry as of the Standard Model, namely $SU(3)\times SU(2)\times U(1)_Y$. Its particle content contains all the elementary particles of the Standard Model and its superpartners. 

Superpartners are corresponding particles to the Standard Model particles, transforming identically under the gauge group symmetries, except their spins are a half-integer different. Every spin-1 gauge boson (photons, W bosons, gluons) has its spin-1/2 gaugino superpartner (photino, Winos, gluinos). Every spin-1/2 matter particles (electrons, neutrinos, quarks, etc.) has its spin-0 scalar superpartner (selectron, sneutrinos, squarks, etc.).

The added complication is that the MSSM cannot have just one Higgs boson, but rather must also possess in the spectrum its gauge symmetry vector complement. There are two key reasons for this. First, unlike in the Standard Model, up and down-like fermions cannot simultaneously get mass from one single Higgs boson within the MSSM. Supersymmetry invariance does not allow the equivalent of a complex conjugate of the Higgs boson giving mass to fermions. Second, the superpartners of the Higgs bosons are spin-1/2 charged fermions, Higgsinos. The Higgsinos, being fermions, contribute to the gauge anomaly. Since the Standard Model fermion spectrum already cancels the anomalies, one cannot add a single Higgsino species without introducing anomalies. The vectorlike complement must too be added, canceling the anomalies as all vectorlike pairs of fermions do. The two reason to have two Higgs boson multiplets within the MSSM are self-reinforcing, rounding out the complete spectrum of the MSSM.

The simplest pure supersymmetric theory has all particle masses equal to their superpartner masses. If this were the case in the MSSM, the selectron would be 0.511 MeV, which would have been seen long ago. Instead, supersymmetry breaking must be present in the spectrum to lift the superpartner masses away from their particle partner masses. Supersymmetry breaking can be accomplished in a way analogous to the spontaneous symmetry breaking of the electroweak symmetry of the Standard Model. This gives rise to an overall superpartner mass scale $\tilde m$. All superpartner masses are of order $\tilde m$ in the simplest approaches to supersymmetry breaking. More complicated structures of supersymmetry breaking will disturb that pattern. For example, the next simplest approach to supersymmetry breaking -- ``charged supersymmetry breaking" -- will have gaugino masses much lighter than scalar masses. This possible large hierarchy between the gauginos and scalar superspartners, and its implication to belief in supersymmetry, will be discussed in more detail below.

Coming back to the simplest form of MSSM: the MSSM is a theory that contains the Standard Model in all ways that one would expect the Standard Model to manifest itself at low energies, but also at a higher scale -- the supersymmetry breaking scale $\tilde m$ -- all the superpartner masses are approximately $\tilde m$ in value. Searching for supersymmetry breaking then becomes an exercise in raising collider energies to high enough values, ${\rm Energy}> 2\tilde m$, such that superpartners are created in the collisions and then are detected directly or indirectly through its decays by the experiment's detectors. 

This exploration strategy of raising collision energy to make superpartners, was what gave rise to hope that when the Large Hadron Collider at CERN turned on, colliding protons at record-breaking energies, supersymmetry would be found. However, no superpartners have been found at LHC at this date. The proton-proton collision energy of the LHC is just under $14\,{\rm TeV}$, which naively would suggest that superpartners have been ruled out up to about $7\, {\rm TeV}$. However, the protons are made of constituent quarks, and the actual collision energy is between quark-quark, quark-gluon and gluon-gluon interactions, which are only a fraction of the energy of the LHC. Furthermore, the production cross-sections to find the superpartners depend on their gauge symmetries. Strongly interacting superpartners are more copiously made than weakly interacting superpartners. And finally, the ability of the detector to see the decay products of the superpartners depends on the details of the detector capabilities and the details of the decay products of the superpartners. 

Combining all the considerations described above one arrives at a complicated array of mass limits for various superpartners. Roughly speaking, all electroweak superpartners (selectron, sneutrinos, charginos, etc.) have been ruled out up to 0.5 TeV, and all strongly interacting superpartners (squarks and gluinos) have been ruled out up to about 3 TeV.  This is much oversimplified, and the details matter when considering precisely defined theories with carefully analyzed data. However, the fact remains that superpartners have not been found up to mass scales significantly higher than previous limits before the LHC.

\subsection{Why SUSY?}

It is a striking aspect of SUSY that a wide range of arguments at various levels speak in favor of its viability. We address the theory's viability rather than its truth to avoid the complications associated with the concept of approximate truth. A theory is viable within a given domain of applicability if the theory's predictions are in agreement with all data that can be collected in that domain (\cite{Dawid2013}). In the context of high energy physics (HEP), the domain of applicability is primarily characterized by an energy scale below which the given theory or model can be viewed as a viable effective theory.\footnote{Effective realism (\cite{Williams2019}), which has been developed specifically for the context of HEP, employs a notion of approximate truth that is similar to the concept of viability used in this paper.}  One can distinguish three types of arguments in favor of SUSY at low energy scales:

\vspace{5mm}

{\bf I: "Esthetic" arguments}

\vspace{2mm}

{\bf 1. Symmetry structure:}	SUSY is the most extensive symmetry involving internal and spacetime degrees of freedom.

{\bf 2. Gravity:}	SUSY provides a natural and appealing basis for integrating gravity within a gauge theory framework (supergravity). 

\vspace{5mm}

{\bf II: Argument from theoretical embedding}

\vspace{2mm}
 
{\bf 3. SUSY is implied by string theory:} Though string theory has not found empirical confirmation, many theoretical physicists think that there are strong reasons to expect that string theory is a viable theory. The epistemic mechanism that leads to this belief has been described in \cite{Dawid2013, Dawid2019} and has been given the name meta-empirical confirmation \cite{Dawid2021}. Meta-empirical confirmation of string theory feeds down to credence in SUSY. String theory does not seem to favor low energy SUSY specifically, however, which makes significance of string theory for low energy SUSY difficult to assess. 

\vspace{5mm}

{\bf III: Empirical arguments} 

\vspace{2mm}

These are arguments based on empirical observations that would be more probable if SUSY was a viable theory. We  list four arguments of this kind that have played a significant role in increasing credence in SUSY. It goes without saying that the listed arguments fall far short of conclusive confirmation of the kind that would be associated with an empirical discovery of SUSY. However, their structure resembles the structure of empirical confirmation: credence is generated by the agreement between a theoretical prediction and empirical data. 
We will frame some of the arguments in a slightly non-standard way to bring out this point. 

\vspace{5mm}

{\bf 4. Dark matter:} Low/moderate energy SUSY with a stable lightest supersymmetric particle (LSP) makes it natural to expect that the LSP plays the role of a significant dark matter component. It therefore significantly increases the probability that galaxy rotation curves and large-scale dynamics of the universe deviate from the observed matter based predictions. This is what is indeed observed.

{\bf 5. Gauge coupling unification:}  SUSY in conjunction with grand unified theories (GUTs) predicts a pattern of coupling constants at the electroweak scale that meet at a plausible GUT scale based on  RG-running that includes SM + low/moderate energy SUSY. This pattern of coupling constants is indeed observed.

{ \bf 6. The value of the Higgs mass:}  The connection between the gauge couplings and the Higgs boson self coupling within supersymmetry predicts a light Higgs boson mass well below twice the $Z$ boson mass. This is a far more restrictive prediction for the Higgs mass than the Standard Model's prediction. The value of the Higgs mass well below $2m_Z$ was indeed observed. 

{\bf 7. The separation of scales:} If we just knew the value of the Planck scale, perturbative corrections from any high energy scalar sector around that scale under normal conditions would render it overwhelmingly likely that the electroweak scale sits close to Planck scale. Low energy SUSY implies the cancellation of quadratic contributions and therefore would substantially increase the probability of a separation between the electroweak and the Planck scale. The observation that the electroweak scale is much lower than the Planck scale therefore increases the credence in low energy SUSY.

\vspace{5mm}
The three groups of arguments work in different ways and at different levels. One important difference is that arguments 1-3 are general arguments in support of SUSY. They do not favor low energy SUSY. Their relevance for low energy SUSY thus is entirely indirect, based on the consideration that finding a phenomenon at a given energy scale becomes more probable once one conditions on the existence of the phenomenon.  
The arguments listed in group III, to the contrary, do speak in favor of SUSY at low, or in the case of argument 4, low or moderate energy scales. 

Assuming that supersymmetry provides the dark matter through its lightest supersymmetric partner (argument 4), the scale of supersymmetry should not be too far from the weak scale. For weak-strength interactions, such as is the case for an LSP, the relic abundance scales upward with the LSP mass. For an LSP mass too high the computed relic abundance is much higher than the inferred dark matter density in the universe. This is equivalent to saying that the universe becomes matter dominated much too early, in conflict with cosmological measurements such as the cosmic microwave background radiation. Given the rather large space of possible parameter values of supersymmetry that feed into this relic abundance calculation, it is not possible to put a rigorous upper bound on all the superpartner masses, but the LSP needs to be under a few TeV for it to be viable dark matter (wino dark matter 2.3 TeV, higgsino dark matter 1.1 TeV, to name two appealing prospects). This is well below the Planck scale, but also well above LHC capabilities to discover.

The measured Higgs mass also constrains the supersymmetry partner masses to not be too high (argument 5), and in particular not to be near the Planck scale. Within supersymmetry the Higgs mass is computable in terms of other parameters in the theory. Most relevant here is that the Higgs mass prediction climbs with higher supersymmetry partner masses of the the top quark, especially. The dependence is only logarithmic, $m^2_h\propto A\ln \tilde m_{\rm top}/m_Z$, where many other mixing angle parameters of supersymmetry and the standard model are embedded in the coefficient $A$. Logarithmic dependence means that small changes in $A$ correspond to large changes of the logarithm's argument in order to keep the result $m_h^2$ the same. For this reason, the scale of supersymmetry breaking is expected to be bounded by the Higgs measurement but without other measurements that scale can be as high as $10^6\gev$ for generic supersymmetric and somewhat higher for non-generic cases (\cite{Ellis:2017erg}).

Gauge coupling unification, like the Higgs boson mass, puts some pressure on supersymmetry to be light, but the dependences are logarithmic on the superpartner masses and therefore not constrained tightly. As will be discussed below, for supersymmetry breaking near the weak scale the three gauge couplings come remarkably close together when they are renormalization group evolved to the scale of about $10^{16}\gev$. The couplings are so close together that it is almost too close compared to generic expectations of a grand unified theory. However, if supersymmetry breaking is at higher scales, such as $10^{6}-10^9\gev$, the threshold corrections required for unification are of the order needed for generic $SU(5)$ and $SO(10)$ supersymmetric grand unified theories. Supersymmetry partner masses above that scale, or no supersymmetry at all, would require threshold corrections at the high scale that are larger than generic expectations from a grand unified theory. Thus, from the point of view of gauge coupling unification within a generic grand unified theory, there is strong pressure to keep supersymmetry masses much lower than the Planck scale but the constraints are not strong enough for LHC results to have much impact on the parameter space of viable unified supersymmetric theories (\cite{Ellis:2017erg}).

\section{Empirical Evidence}

\subsection{Empirical Evidence for Grand Unified Theories}

\noindent
{\it The meeting of gauge couplings (evidence $E_1$)}

Long ago it was noticed that the matter content (i.e., the fermions) of the SM could be arranged into complete multiplets of a higher dimensional gauge group. For example, $(d_L, u_L, u_R, e_R)$ have all the appropriate charges to be embedded in a complete ten dimensional representation of $SU(5)$ (${\bf 10}$), and $(e_L,\nu_L,d_R)$ can be embedded in a complete five-dimensional representation of $SU(5)$ ($\overline{\bf 5}$). This gave rise to the notion that the three gauge groups of the SM are really just remnants of a spontaneously broken grand unified theory's (GUT), such as $SU(5)$. 

Other GUTs are possible, including $SO(10)$, if one considers the possibility of a right-handed neutrino. The latter is widely assumed given the realization that neutrinos have mass, albeit very small.  The appealing aspect of $SO(10)$ is that all fermions can be embedded into a complete sixteen dimensional representation (${\bf 16}$). The dimensionality can be understood by realizing that each quark has three colors, and thus $u_L$, $d_L$, $u_R$ and $d_R$ have a total of 12 degrees of freedom. One then adds the degrees of freedom associated with $e_L$, $e_R$, $\nu_L$ and $\nu_R$ to reach the total of 16.

The bosons of the SM, on the other hand, do not on their own constitute complete representations of the SM. But this is fine for the viability of the GUT for two reasons. First, whenever there is spontaneous symmetry breaking of a gauge symmetry it generally leads to a mass gap for bosons. In other words, vector bosons and gauge bosons get mass. This happens within the SM itself when $SU(2)\times U(1)$ gauge group spontaneously breaks down to $U(1)$ giving rise to massive $W^\pm$ and $Z$ bosons and a massive Higgs boson at the scale of that symmetry breaking. For a GUT the symmetry breaking scale would be much higher -- perhaps near $10^{16}\, {\rm GeV}$ for reasons discussed below -- which would give mass to a collection of vector bosons and Higgs bosons that are much higher than any experiment could directly probed. 

The massive states of spontaneous symmetry breaking of the GUT are the missing states to round out a complete multiplet. For example, in $SU(5)$ the vector bosons are in a 24-dimensional representation whereas the SM gauge bosons (gluons, $W^\pm$ and $Z$ bosons, and photon) are only 12 of those degrees of freedom, leaving 12 exotic $X,Y$ bosons at a very high mass scale. Likewise the Higgs boson of the SM is two degrees of freedom in a 5-dimensional representation of $SU(5)$ leaving three degrees of freedom unaccounted for. Those three degrees of freedom is the so-called ``color triplet Higgs" which has a mass near the GUT-symmetry breaking scale. 

The scale of spontaneous symmetry breaking for the GUT gauge group is not known, but it should be roughly near the scale of the unification of the gauge couplings associated with the electromagnetic, weak and strong forces. After normalizing the abelian gauge group coupling appropriately to take into account proper group theory factors, the couplings have their closest approach to unifying at around $10^{14}\,{\rm GeV}$ in a non-supersymmetric theory. For supersymmetric theories possessing superpartner masses near the TeV scale, the unification scale is approximately $10^{16}\,{\rm GeV}$. (For a comprehensive review of Supersymmetric GUTs, see \cite{Raby:2017ucc}.)

From our infrared perspective the gauge couplings should not unify perfectly when we run them up to their higher energy scale values using renormalization group evolution. One expects small deviations due to high scale threshold corrections. In other words, those higher mass states that obtained mass associated with the spontaneous symmetry breaking will contribute to quantum corrections at those high energies that will disrupt the gauge couplings from taking on perfectly unified values given that the representations are ``broken" from the symmetry breaking. 

The near perfect unification of the gauge couplings when run up to higher scales within a supersymmetric framework is therefore somewhat surprising given the expected threshold corrects at the GUT scale. The tiny value of the mismatch is not so small that it cannot be accommodated by reasonable threshold corrections, but it does gives rise to some speculations that the threshold corrections at the high scale are so small that it cries out for ``precision unification" scenarios that suppress threshold corrections more than would be ordinarily expected in a straightforward GUT theory. 

The mismatch of the gauge couplings within a non-supersymmetric assumption has two problems compared to the supersymmetric theory. First, the mismatch is uncomfortably large, requiring threshold quantum corrections from GUT-scale states to have values somewhat larger than would be naively expected out of a GUT. The second worry is that the naive value of unification in the non-supersymmetric scenario occurs at too low a scale to be phenomenologically viable. The issue is that low unification scale means that the GUT-scale exotic states have masses at that low scale too, and they will generically induce proton decay at a rate faster than current experiment says is allowed. We think that criticism of non-supersymmetric unification is not independent of the question of uncomfortably large threshold corrections. As long as one accepts large threshold corrections it is not a problem to have unification happen at a higher scale, say $10^{16}\, {\rm GeV}$, with its accompanying large threshold corrections to realize it. That in turn lifts the GUT-scale masses to a sufficiently high value that they push the proton decay lifetime to larger values, escaping the experimental limit.

\medskip
\noindent
{\it Unification of gauge coupling constants (evidence $E_2$)}

We wish to estimate the probability that randomly selected gauge couplings would have values accidentally consistent with grand unification if grand unification is not an underlying fact of nature. This of course is an impossible question to answer precisely. However, one does have the intuition, as some physicists have said, that it would be a cruel joke of nature if there is no grand unification given how unified the couplings appear to be. We attempt to build on that intuition here.

First, by ``appearing unified" we mean that there is a renormalization group scale $\mu$ -- the unification scale -- where all three gauge couplings are very close. And by ``very close" we mean that reasonable size threshold corrections that arise from remnant states from grand unification breaking are large enough to account for the small differences between the gauge couplings, as discussed above. The relevant equations to be satisfied for unification are
\begin{equation}
\frac{1}{g^2_i(\mu)}-\frac{1}{g^2_j(\mu)}=\frac{\Delta\lambda_{ij}}{48\pi^2},~~i,j=1,2,3
\end{equation}
where $g_i(\mu)$ are the gauge couplings at the unification scale $\mu$, and $\Delta\lambda_{ij}$ characterize the threshold corrections at the high scale associated with remnant states charged under the $i$ and $j$ gauge groups. 

To test for possible unification we must ask ourselves what is the maximum value of $\Delta\lambda_{ij}$ that a unified group would allow? Within supersymmetric MSSM theory the maximum $\Delta\lambda_{ij}$ needed is less than 10, which would constitute very small high scale threshold corrections. On the other hand, the maximum the Standard Model would require is about 200, which is a somewhat uncomfortably large value of $\Delta\lambda_{ij}$. (See fig.~2 of \cite{Ellis:2015jwa}.) Many physicists will say that unification is possible within supersymmetry but not within the Standard Model given the measured gauge couplings, and so we will take $\Delta\lambda^{\rm max}_{ij}$ to be 200. Thus, our condition for possible unification is to determine whether this condition is satisfied for some scale $0<\mu<M_{\rm Planck}$:

\begin{equation}
\left| \frac{1}{g^2_i(\mu)}-\frac{1}{g^2_j(\mu)}\right| < 0.42,~~{\rm for}~i,j=1,2,3.
\label{eq:unify_conditions}
\end{equation}
\vspace{5mm}

To obtain a probability for accidentally achieving the condition of unified couplings as expressed by eq.~\ref{eq:unify_conditions} we must first choose a metric over which the gauge couplings are randomly sampled. Of course no metric is technically defensible, and thus we will do the simplest thing and sample on a flat distribution of the coefficient $1/g^2$ of the kinetic term of the gauge fields. This is the standard geometric normalization where the gauge coupling is not in the covariant derivative but rather in front of the kinetic terms. We assume that it is sampled over the range of 0 to $(1/g^{2})_{\rm max}$.

What choice to make for $(1/g^{2})_{\rm max}$? The larger the value chosen the smaller the probability that one will obtain for possible-unification when randomly samplying over $1/g_i^{2}$ and testing for unification using eq.~\ref{eq:unify_conditions}. One should note that within supersymmetry $1/g_i^2=2.0$ and within the Standard Model $1/g_i^2=3.2$. We need to be sure to choose $(1/g^{2})_{\rm max}$ that is at least as high as 3.2. For simplicity we will use $10$ since it is less than an order of magnitude different than $(1/g^{2})_{\rm max}$ and conforms with the standard intuitions often stated among physicists that dimensionless coefficients should be within an order of magnitude of one unless some deeper explanation is at play.  With this choice of $(1/g^{2})_{\rm max}=10$ one finds that $0.5\%$ of randomly selected gauge couplings sampled on flat distrubiton of $1/g^2(\mu)$ have the potential of being unified into a grand unified group according to the condition of eq.~\ref{eq:unify_conditions}. If we had selected $(1/g^{2})_{\rm max}=3.2$, which is equal to the value of the SM at its putative unification scale, one obtains $4.7\%$ of randomly selected cases have potential unification. We can think of no good reason to argue for a higher prospect than this $\sim 5\%$ value; however, we do not expect the Standard Model case to be the edge case. That is why our best estimate is that the probability of three randomly selected gauge couplings unifying is less than $0.5\%$ consistent with $(1/g^{2})_{\rm max}=10$ or higher.

\medskip
\noindent
{\it Uniqueness of GUTs}

As we discussed above, a key motivating factor for considering GUT theories is that all the fermions are unified into complete GUT multiplets. In order to complete that argument we must ask whether there are restrictions on quantum field theories that will also arrange fermions into higher rank groups even if unification does not actually occur. In other words, perhaps the $SU(5)$ and $SO(10)$ symmetries are just accidental organizing symmetries for the fermions and not realized as operational symmetries with their corresponding extra gauge bosons in a unified GUT gauge group.

One option is that there are enforced global symmetries that require the higher symmetry group multiplets. However, there are strong arguments against the existence of global symmetries that enforce restrictions on quantum field theories. Rather, global symmetries are generally thought to only be accidental due to what is made possible in the lowest order terms of an effective theory with a particular gauge symmetry and restricted set of fields with mass below the cutoff of that effective theory. Autonomous global symmetry enforced on a theory are not expected to be at play, and even if they were they would constitute a curious new beyond-the-SM principle to be explored more fully.

Another option is that the requirement the theory has no anomalies generally results in low-energy fermion multiplets that can be organized into representations of simpler symmetry groups. Indeed, there are no anomalies within $SO(n)$ theories.  The conjecture might be considered that if nature constructs a complicated anomaly-free theory with direct gauge product group like $SU(3)\times SU(2)\times U(1)$, where each subgroup has no automatic anomaly cancellation in the presence of fermions, that the final fermion particle content will necessarily arrange itself into something like representations of anomaly free $SO(10)$. This conjecture, however, is incorrect.

One approach to show that the conjecture is incorrect is to immediately recognize that any vectorlike complement of states is anomaly free. In other words, one could construct an entire theory based on a the right-handed up quark and its vectorlike complement. That theory is anomaly free and cannot be embedded in a complete GUT multiplet, and so the conjecture is immediately refuted.

One could refine the conjecture and assume that all vectorlike states get large (gauge invariant) vectorlike masses and the issue is whether the remaining low-energy chiral states must necessarily arrange themselves into simple GUT representations if they are to be an anomaly free set. This conjecture is also false.

For example, if we assume a fermionic particle content identical to the Standard Model's in every way except hypercharge we can get a sense for how unlikely a random spectrum of fermions can be put into a grand unified representation. In this scenario we randomly select integer hypercharge values between -6 and 6 for each of the fermions\footnote{In this normalization, all hypercharges are integers. For example, in the Standard Model the right-handed electron would have hypercharge $-6$, and the left-handed quarks would have hypercharge $+1$.} and retain only unique sets of charges that render the theory anomaly free. We get 62 such cases. Of those 62 cases only 2 have a full complement of charges that enable embedding them into complete representations of $SU(5)$ or $SO(10)$. That gives odds of $1/31=3.2\%$ for random dart throw in hypercharge parameter space, under these assumptions, yielding a ``GUT-able" spectrum. If we allow the random selection of hypercharge to be $\pm Y_{\rm max}$ with $Y_{\rm max}>6$ the number of anomaly free spectra increases by $Y_{\rm max}^2$ at a faster pace than GUT-able spectra. If we added particle content and $SU(3)$ and $SU(2)$ charges among the list of free parameters, the odds of GUT-able anomaly-free spectra go down even further. Thus, we expect the odds to be worse than 3.2\% for randomly chosen spectra being GUT-able. Indeed, a recent preprint, \cite{Herms:2024krp}, has independently suggested ${\cal O}(1/100)$, consistent with our estimate. This gives some leverage to the claim that the GUT-able spectrum of the Standard Model fermions gives positive support to the claims of an underlying GUT theory.

\subsection{Empirical evidence for SUSY}

\noindent
{\it Separation of scales (evidence $E_3$)}

Separations of scale in nature are sometimes naturally understood and other times mysterious. In the first category, few are alarmed that the scale of the proton mass and QCD dynamics is roughly at $\Lambda_{\rm QCD}$, which is several orders of magnitude below the weak scale. The reason is that $\Lambda_{QCD}$ arises from an infrared blow-up of the QCD gauge coupling at $\sim\Lambda_{\rm QCD}$. In other words, it's a quantum induced scale and not a scale directly in the lagrangian. 

In contrast, the separation of scales between the Planck scale and the weak scale is more mysterious. Both appear to show up directly in the lagrangian -- the Planck scale in the gravitational lagrangian and the weak scale in the Higgs boson lagrangian. The large separation of scales is considered by some a mystery that requires explanation, in the sense that this large ratio of scales can be understood by a deeper principle than is known today. Neither the Standard Model nor SUSY have inherent within them an explanation. However, SUSY renders such a separation of scales more likely due to its quantum corrections being controlled by the scale of supersymmetry breaking rather than controlled by the heaviest masses in the theory, like the Standard Model suffers due to its quadratic sensitivity to higher mass scales.

We will account for the lack of full prediction on the side of SUSY, somewhat crudely, by using a logarithmic measure to partition parameter space up to the observed separation of scales. Therefore, we will attribute a probability of 1/16 to the $10^{-16}$ separation between the Planck scale and the electroweak scale.  In comparison, we would say that a theory that predicted the order of magnitude of the observed separation of scales would assign probability one to a separation of scales of that order of magnitude.

\medskip

\noindent
{\it The Higgs mass lies below 150 GeV (evidence $E_5$)}

Up to this point, our modeling has relied entirely on data available before the discovery of the Higgs particle. 
Let us now model a probabilistic analysis of the measured Higgs mass value in light of the SM and low energy SUSY respectively. 

\bigskip\noindent

Recall that within the SM the Higgs boson mass is computed to be $m^2_h=2\lambda v^2$, where $v$ is the Higgs vacuum expectation value, which is determined by the weak boson masses, and $\lambda$ is the Higgs boson self coupling $\lambda$, which is a free parameter.  There are no {\it a priori} constraints on $\lambda$. If the theory is perturbative, in the sense that observables involving multi-Higgs boson scattering are perturbative, then $\lambda$ is constrained to have values equivalent constraining $m_h$ to be less than approximately $700\gev$. There is no other theoretical constraint or hint as to what the Higgs mass bounds might be within the SM context. If the SM is embedded in a perturbative GUT then there is pressure for the Higgs mass to be relatively light, below $\sim 190\,{\rm GeV}$ in order for the Higgs boson self-coupling to remain perturbative up to the unification scale. That enhanced likelihood for the SM-GUT Higgs boson to be light will be reflected in our likelihood tables below. 

Within the supersymmetric context, the Higgs mass is a computable output of the theory once the rest of the spectrum is determined. As discussed above that complication is a complex one, involving many supersymmetric parameters. However, the Higgs mass has a fairly stable upper bound of about $150\gev$ over even extreme values of supersymmetry breaking that push supersymmetry right to the edge of viability for dark matter or viability for gauge coupling unification without requiring unexpectedly large threshold corrections at the unification scale. The fact that the Higgs boson was discovered at a mass value less than $150\gev$ is completely expected within the supersymmetric context. In the SM context there is nothing {\it a priori} special about such a low Higgs mass compared to a Higgs mass hundreds of GeV higher. This asymmetry in expectations will drive some of the Bayesian results below.

In light of the discussion in Section 3.2, we will use parameter space linearly as probability space in the upcoming analysis. In other words, if a hypthesis $H$ allows for a parameter range $R$ for a given variable, the probability that $H$ is viable and the parameter assumes a value within the smaller range $R_1$ is assumed to be $P(H_{R_1})=P(H)\frac{R_1}{R}$.

It must be noted that, apart from SUSY, there are other scenarios that also put strong limits on the Higgs mass. Constituent Higgs models, (which seem difficult to sustain in light of current Higgs data, though), would just put a slightly higher limit on the Higgs mass than low energy SUSY, as would the requirement that the SM survives all the way up to the Planck scale. Other scenarios, such as asymptotic safety with mild beyond-the-SM corrections or  the scenario of a metastable Higgs potential arguably lead to stronger constraints on the Higgs mass than low energy SUSY. Since the latter scenarios are not incompatible with low energy SUSY, increased credence in them would not translate into correspondingly decreased credence in low energy SUSY, though there would be some detrimental effect on SUSY credences.

\subsection{Empirical evidence that lowers credence SUSY}

\noindent
{\it The failure to measure Proton decay (evidence $E_4$)}

\bigskip\noindent
 
Proton decay has a rich history of enthusiastic expectations followed by disappointments. When Grand Unified Theories were first postulated it soon became apparent that one of its most important implications was the prediction of proton decay. Proton decay becomes allowed by exchange of heavy $X$, $Y$ bosons that are remnants of the spontaneous breaking of the grand unified gauge group down to the Standard Model gauge group: $G_{\rm GUT}\to SU(3)_c\times SU(2)_L\times U(1)_Y$. As discussed above, in the case of $G_{\rm GUT}$ being $SU(5)$ there are 24 gauge bosons of this grand unified group and $8+3+1=12$ gauge bosons of the Standard Model, leaving 12 heavy bosons obtaining mass from the GUT breaking mechanism $M\simeq M_{\rm GUT}$. The lifetime of the proton is computable within the GUT scenario, which scales like $\tau\sim M_{\rm GUT}^4/m_{\rm proton}^5$. In addition to the lifetime of the proton being highly sensitive to the precise value of the GUT scale, the  proportionality constant in the lifetime calculation is a subtle and difficult calculation. 

Much work went into the prediction of the proton's lifetime if grand unifications was indeed correct. The theoretical efforts culminated in a comprehensive review article by~\cite{Langacker:1980js}. The general expectation was that the proton should decay within about $10^{30}\,$ seconds. Experiments were designed to reach this lifetime target. In summer of 1983 it was announced by the Irvine-Michigan-Brookhaven (IMB) collaboration that their search for proton decay was unsuccessful and declared at 90\% confidence level a lower bound on its lifetime at $1.9\times 10^{31}\,{\rm sec}$. 

The IMB announcement of its failure to measure proton decay was a surprise and an extreme disappointment (see~\cite{Georgi:1983}).
The excitement surrounding GUTs waned because there was so much confidence that the experiments {\it should have found} the proton to decay. However, as is explained in~\cite{Ellis:2015jwa} the theory confidence was unjustified that they had bounded well the expectations for proton decay. In our modern understanding of effective theories and matching conditions, the uncertainties in the original calculations were much too low. Indeed, even within the Standard Model (i.e., non-supersymmetric context, see below) it was not justified to have such a high confidence that IMB would find proton decay. Indeed, as~\cite{Lavoura:1993su} pointed out a decade later, a proton lifetime significantly above the IMB limits is straightforward to contemplate in $SO(10)$ grand unified theories, and in fact is consistent even with the improved present-day proton decay lifetime limits.

Such understanding was not appreciated in the mid- to late-1980s, and the confidence in GUT theories remained low. However, the precision measurements of the new LEP collider that measured $e^+e^-$ collisions on the $Z$-boson mass pole changed attitudes dramatically in the early 1990s. \cite{Amaldi:1991cn} and others took the precision measurements that were being made at LEP and extrapolated the gauge couplings to the high scale, where they noticed that the gauge couplings unified to an astonishingly good fit at a higher GUT scale within an assumed supersymmetric framework. The calculated unification scale was raised from approximately $10^{13}-10^{14}\,{\rm GeV}$ to $2\times 10^{16}\,{\rm GeV}$ in this modern formulation. Since the proton decay lifetime scales as the fourth power of the GUT scale, $\tau_p\propto M_{\rm GUT}^4$, the corresponding prediction for the lifetime would be significantly higher than the IMB and Superkamiokande limits that had been found in the 1980s. 

The excitement surrounding GUTs was reinvigorated with these new precision measurements and calculations, and a significant effort was initiated to explore all aspects of supersymmetric grand unified theories. Many hundreds, if not thousands, of research articles were written on GUT theories, ranging from experimental predictions of proton decay to possible patterns for the neutrino masses coming from a GUT structure. Viable theories were obtained and excitement remained high.

None of the improved limits on proton decay created a crisis in confidence in supersymmetric GUTs. There was hope that we would be lucky and that the SUSY GUT theories on the low end of the proton decay lifetime prediction would be the one that nature chose so that we could see proton decay from those experiment. However, the lack of evidence of proton decay did not materially affect confidence in SUSY GUTs because the range of proton lifetime prediction was so large, extending to well beyond the current experimental limits, or even the projected limits of the next generation proton decay detectors (see, e.g.,~\cite{Bhattiprolu:2022xhm}). Instead, confidence in SUSY GUTs has been shaken more by the lack of discovery of superpartners than in the lack of discovery of proton decay, for reasons we will discuss below.

\medskip

\noindent
{\it The failure to find SUSY signatures at LEP2 and the LHC (evidence $E_6$)}

\bigskip\noindent

The symmetry of supersymmetry cannot be fully realized in nature, otherwise scalar electrons would have the same mass as ordinary electrons, etc., which is in gross violation of experimental limits. Thus, if supersymmetry is correct it needs to be broken. This is no drawback on the theory, since many of our favorite symmetries in nature are spontaneously broken, including chiral symmetry and electroweak symmetry. Without knowledge of the superpartner masses there is no way to know at what scale supersymmetry is spontaneously broken. Thus, it remains a free parameter. As discussed above, this free parameter that sets the general scale, up to factors of ${\cal O}(1)$, is sometimes denoted $\tilde m$. In principle it can be anywhere from $\tilde m=0$ to $\tilde m=M_{\rm Pl}\simeq 10^{18}\,{\rm GeV}$. There is no firm data guide to tell us where we should anticipate $\tilde m$ to be, except above  current experimental limits.

On the other hand, there is a philosophical guide, in a sense, that has convinced many that $\tilde m$ should not be significantly larger than the electroweak scale. In other words, $\tilde m \gg m_{Z}$ is ``unlikely" in the minds of many physicists. The reason is that the electroweak potential that gives rise to the $Z$ boson mass in a supersymmetric theory is comprised of several particles with mass $\tilde m_{i}\sim \tilde m$ that must conspire within the theory to give the $Z$ mass. Roughly speaking, spontaneously broken supersymmetric theories yield a condition of $\tilde m^2_1-\tilde m^2_2=m_Z^2$. If $\tilde m_{1},\tilde m_2\gg M_Z$ it would be an extreme finetuning for this condition to be satisfied. In other words, it is unexpected that two mostly uncorrelated very large numbers ($\tilde m^2_1$ and $\tilde m^2_2$) should cancel each other out to yield a very tiny number in comparison ($m^2_Z$).

The LHC has presently set bounds on the masses of electroweak superpartner masses of approximately ${\rm few}\,{\rm TeV}$. The finetuning in the electroweak potential then appears to be approximately one part in $\tilde m^2/m_Z^2\sim 10^3$. There are bickerings about precise definitions of finetuning, and the relative merits of precise spectra of supersymmetric theories to reduce the finetuning in the face of the current LHC constraints, but generally speaking it is hard to get around the worry that a supersymmetry theory is finetuned to at least one part in a hundred.

This worry is in the philosophical domain, because there is no experiment or mathematical consistency condition that breaks down if a theory is finetuned at this level, or any level for that matter (see, e.g.,~\cite{Giudice:2008bi,Wells:2018sus,Craig:2022eqo}). In order to turn it into a rigorous probabilistic argument for the finetuning to be unlikely one needs to assume the parameters of the theory are random variables selected by some mechanism in the early universe, and one must also assume a measure on the parameters of the theory. No precise assumptions of this nature can be justified as incontrovertible. 

As a result, the community of physicist roughly splits into two groups. The larger group considers finetuning a real issue and agrees with the sentiment that a highly finetuned manifestation of a theory is less expected than a non-finetuned manifestation of a theory. Where to draw the line on what is unacceptably large finetuning in a theory is murky, for reasons stated above. However, most feel some level of discomfort, differing from person to person, in supersymmetric theories given the level of finetuning that it has after not having been discovered at the LHC. A smaller group if physicists reject this way of thinking and consider finetuning a non-issue because it cannot be rigorously analyzed and therefore should not be in the vocabulary of a physicist and most certainly should not be included in judging criteria for or against a theory's viability. For members of this group the lack of discovery of supersymmetry induces no reduction, or at least should not induce any reduction, in belief in the theory since superpartner masses could be at any scale all the way up to the Planck scale of $10^{18}\, {\rm GeV}$, many orders of magnitude beyond current experiment.

\section{A Bayesian analysis of SUSY's prospects}

\subsection{Setting the stage}

The present paper aims at understanding the significance of arguments in support of low energy SUSY from a Bayesian perspective. The arguments spelled out in the previous section are not all equally suitable for a Bayesian representation. 

The aesthetic arguments 1 and 2 seem difficult to model in terms of Bayesian updating at all. Argument 3 hinges on credence in string theory, which has not found empirical confirmation. As described in \cite{Dawid2019}, the generation of such credence can be modeled in a Bayesian framework as \textit{meta-empirical theory confirmation}, which requires admitting evidence that lies outside the theory's intended domain. In this paper, however, we will focus on providing a Bayesian model of \textit{empirical} confirmation, that is, on confirmation by observations in the theory's intended domain. Arguments 4-7 constitute empirical confirmation. Since, as we will briefly discuss below, argument 4 involves some intricacies that render Bayesian modeling difficult, we will provide a Bayesian model of updating under evidence only for arguments 5-7.  

The fact that Bayesian modeling is more complicated for arguments 3-4 and difficult to conceptualize at all for arguments 1-2 does not imply that those arguments are insignificant. They are taken seriously by many physicists and thus do have an impact on the credence in low energy SUSY. In our discussion, we will treat all four arguments as contributing to what will be the informed prior probability we will use as a starting point for the full Bayesian analysis.

\subsection{Bayesian Epistemology}

Bayesian epistemology\footnote{For general presentations of Bayesian epistemology, see \cite{Urbach1993-HOWSRT} and \cite{Sprenger2019-SPRBPO}} is based on the understanding that confidence in a theory's viability can be expressed in probabilistic terms.
The change of credence in a theory follows Bayes' formula:

\begin{equation}
P(H|E) = \frac{P(E|H)}{P(E)}P(H) \label{Bayes1}
\end{equation}

$P(E)$ can be written as the total probability
\begin{equation}
P(E) = \sum_{n=1}^{N} P(E|H_n)P(H_n)  \label{Bayes2}
\end{equation}
where $n$ runs over all possible theories that can account for the available data. This includes known alternatives as well as unconceived alternatives, the contribution of which is often modeled in the form of the catchall hypothesis $H_{CA}$: "The true/viable theory is not among the known theories".

Evidence $E$ confirms a theory $H$ if  
\begin{equation}
P(H|E)>P(H)   \label{Bayes3}
\end{equation}
Canonically, $P(H)$ would be called the probability that hypothesis $H$ is true.
Since we are interested in absolute credences, however, we need to be careful to spell out precisely what it is physicists have credence in when they endorse a theory in high energy physics. The theories whose status we aim to assess are deployed as effective theories: if endorsed, an effective theory is expected to adequately represent the data up to a given energy scale. Whether or not it would  make sense to call such a theory approximately true is a philosophical question in the the context of the scientific realism debate, but is of little concern for the issues discussed in this paper. We therefore avoid using the notion of truth and define $P(H)$ as the probability that theory $H$ is \textit{viable} with a given precision up to a given energy scale (as introduced in Section 2).
Strictly speaking, $P(H)$ would thus be indexed by an energy scale. For the sake of simplicity, we will omit that index whenever it is not relevant for our analysis.

In order to understand the nature of our analysis, it is helpful to compare the project of Bayesian epistemology to Bayesian data analysis. In recent decades, there has been an upsurge of the use of Bayesian methods in physics. Bayesian methods are employed in order to account for physical background knowledge that influences the interpretation of incoming data.
The use of Bayesian methods in physical data analysis is guided by one core principle: the specification of priors should, to the extent possible, account for objectifiable physical background knowledge without admitting subjective bias in favor or against any of the models, theories or hypotheses that are to be compared to each other based on the available data.\footnote{To be sure, physicists are not fully successful in achieving that goal. Much debate in Bayesian cosmological data analysis circles around choosing the most adequate choice of priors.} 
Two main strategies are deployed to achieve that goal. First, one aims to attribute generic priors to parameter values that are allowed by physical background knowledge. Second, one often focuses on comparative data analysis, which ejects $P(E)$ from the analysis based on the relation:

\begin{equation}
 \frac{P(H_1|E)}{P(H_2|E)} = \frac{P(E|H_1)P(H_1)}{P(E|H_2)P(H_2)}   \label{Bayes4}
\end{equation}

\vspace{5mm}

These two strategies aim to bring Bayesian data analysis as close as possible to the extraction of objectifiable numerical results. It does so by disregarding, to the extent possible, personal assessments of a theory's prospects, and by avoiding the specification of absolute credences in individual hypotheses. 

Bayesian epistemology in the context to be addressed in this paper is driven by the opposite agenda. The aim is to model the physicist's expectations regarding the prospects of a given, so far empirically unconfirmed, theory.
To develop a Bayesian model of this kind, we need to address absolute credence and account for the impact of data on the subjective choice of priors scientific Bayesian data analysis tries to disregard.  

\subsection{The probabilistic significance of parameter space}

Empirical confirmation in a Bayesian sense is based on one core characteristic: credence in theory gets increased by the fact that the theory favors a narrower patch of parameter space around the measured values than (most of) its contenders. On a Bayesian account, this generates confirmation if one treats the theory as a set of "sub-theories" representing the individual parameter values. In the absence of further information, equal credence would then be given to each of those "sub-theories", which would amount to translating parameter space into probability space. This step faces well-known issues associated with choosing a probability measure in the absence of a genuine physical dynamics for selecting a parameter value. On the one hand, parameter space has often been viewed uncritically in terms of probability space in order to establish the significance of fine-tuning arguments in particular. And yet, it has been argued by some at various levels of intensity  that no probabilistic conclusions should be drawn at all from constellations that play out in parameter space as long as such probabilistic implications cannot be deduced from physical laws. The literature ranges from the most extreme versions of this argument (\cite{Hossenfelder:2018ikr}) and the fairly skeptical perspective of \cite{Grinbaum2012-GRIWFA}, to an exploration of the problematic implications of having an extreme skepticism of finetuning arguments (\cite{Wells:2018sus}), to a focused technical discussion within String/M-theory framework (\cite{Douglas:2019kus}).

The discussion in the present paper will be based on the following considerations:

\vspace{5mm}

{\bf I. Bayesian epistemology formally relies on assigning probabilities to alternative hypotheses.} As discussed above, specifying $P(E)$ includes the implicit specification of probabilities of known and unconceived alternatives to the hypothesis $H$ under scrutiny. Refraining from such implicit specifications would amount to leaving P(E) entirely undetermined, which would block any significant confirmation of $H$ based on updating on $E$.\footnote{We write "significant confirmation" since formal proofs of incremental confirmation may get through based on ruling out $P(H_i)=0$ for any hypothesis consistent with the data.} 

{\bf II. Reading parameter space probabilistically is one specific way of implementing the requirement spelled out in point I.} In light of the reasoning in point I, it seems inconsistent to exclude a probabilistic view on parameter space in the context of a Bayesian analysis.

{\bf III. The measure problem indicates uncertainties regarding the specific way to interpret parameter space probabilistically.} A naive translation of parameter space into probability without physical motivation thus can serve a starting point for Bayesian analysis but has limited epistemic significance. 

{\bf IV. Expectations with regard to so far hypothetical or unconceived fundamental physics are important.} Probabilistic readings of parameter space become epistemically more powerful when they are supported by expectations regarding more fundamental theories that provide a dynamical explanation of the selection of parameter values.

\vspace{5mm}

In practice, probabilistic considerations about parameter space tend to be informed by tentative views on possible underlying physical reasons for selecting a certain measure. For example, the specification of low energy parameter values may be understood in terms of the dynamical selection of a ground state of string theory. In some cases, such as naturalness, doubts about the existence of a fundamental physical basis for selecting a probability measure has lead some observers to questioning the significance of probabilistic reasoning in the given context. In our analysis, we will not spell out that background reasoning for a probabilistic understanding of parameter space in detail. We will assume that it is acknowledged in conventional cases. This attitude seems to be shared by supporters of SUSY who emphasize the significance of SUSY's consistency with the measured Higgs mass, but also by SUSY skeptics who emphasize the significance of the fact that low energy SUSY has not been found after having excluded much of its parameter space. In the specific case of naturalness, where the legitimacy of probabilistic reasoning in the given context has been a matter of debate, we allow for an additional factor controlling the argument's significance.

\subsection{Explaining the quantitative approach}

Our analysis will play out at two levels. First, we aim at a better understanding of the way credences in SUSY are generated. At this level, priors are merely placeholders used for representing the general mechanisms of updating under relevant data.
Our second aim is to understand how actual credences in SUSY can be traced back to the selections of priors. This involves understanding inter-dependencies of prior selections and posteriors that can be plausibly reached based on specific general attitudes towards theories' prospects. In that discussion, actual priors do play a role.

We will explicitly spell out the credences of various sample agents, which represent specific general attitudes towards theory assessment. Those quantitative analyses aim to convey a rough intuition on the ways in which actual credences, to the extent they are rationally motivated, can be construed as representing attitudes towards theory assessment.
The three core stances we will analyze are the following. First, in order to introduce the Bayesian representation of arguments in support of SUSY, we will take the stance of an optimistic exponent of SUSY, to be called agent A, who attributes considerable weight to naturalness considerations. Optimism, in our Bayesian representation, will amount to being somewhat confident that scientists who have worked extensively on a given subject matter and think that they have a good understanding of the issues, have found most of the promising approaches towards the problem they are dealing with. In other words, the optimist does not by default expect that an unconceived alternative will be needed to explain the observed phenomenology in a given domain.  

In comparison, we will extract credences of an agent C who is fairly skeptical about scientists' capacity to assess a theory's prospects of being viable.
In between, we will model a "neutral" agent B. 
In addition, we will consider an agent B' who shares the views of our neutral agent B in general but is skeptical about naturalness arguments.

\section{The model}

\subsection{Evidence and Hypotheses}

\vspace{3mm}
\subsubsection{Empirical Evidence}
\vspace{3mm}

We define low energy SUSY as SUSY that can be empirically established below a given energy scale, {such as the TeV scale or the PeV scale in the case of split supersymmetry.}
We want to update credence in low energy SUSY under several pieces of empirical signatures $E_i$, as discussed in Section 3:

\vspace{3mm}

$E_1$: observed particles fit in GUT representations.

\vspace{3mm}

$E_2$: RG running of the three gauge couplings from their observed low energy values show 

convergence at a high energy scale (the GUT scale).

\vspace{3mm}

$E_3$: The electroweak scale and the Planck scale are separated by 16 orders of magnitude.

\vspace{3mm}

$E_4$: No proton decay has been observed.

\vspace{3mm}

$E_5$: The Higgs mass is below 150 GeV.

\vspace{3mm}

$E_6$: No signatures that point at SUSY have been found at LEP2 and the LHC.

\vspace{5mm}

\vspace{3mm}
\subsubsection{The Basic Hypotheses}
\vspace{3mm}
 
The conceptual basis for all theoretical approaches discussed is provided by the empirically confirmed standard model of particle physics (SM). Low energy SUSY is a theory that reaches beyond the standard model.
Besides low energy SUSY, we consider GUT (Grand Unified Theory) as a second theory BSM beyond the SM. We thus partition the space of possible theories into four basic hypotheses $H_k$:

\vspace{5mm}

$H_{SM}$: the world shows no low energy SUSY and no GUT.

\vspace{3mm}

$H_{SUSY}$: the world shows low energy SUSY and no GUT.

\vspace{3mm}

$H_{GUT}$: the world shows GUT and no low energy SUSY.

\vspace{3mm}

$H_{SUSYGUT}$: the world shows low energy SUSY and GUT.

\vspace{5mm}

On our definition, hypotheses 1-4 exhaust the space of possibilities. We therefore have:

\begin{equation}
P(H_{SM}) + P(H_{SUSY}) + P(H_{GUT}) + P(H_{SUSYGUT}) = 1
\end{equation}

\vspace{5mm}

\vspace{3mm}
\subsubsection{Further Hypotheses}
\vspace{3mm}

Any new physical concepts that reach beyond SUSY and GUT are treated as specifications of one of the four basic hypotheses.
Due to the role of total probability in Bayes' formula, updating the credence in a hypothesis $H$ under evidence $E_i$ that is in agreement with $H$ crucially depends on the prior credence in alternative hypotheses that are also in agreement with $E_i$. Therefore, we need to model the spectrum of possible alternative explanations. We introduce, for each piece of empirical evidence $E_i$ that can be explained by SUSY, GUT, or SUSYGUT, a hypothesis $U_i$:

\vspace{5mm}
$U_i$: Empirical evidence $E_i$ can be explained by a viable physical theory beyond the SM that is neither SUSY nor GUT. 
\vspace{5mm}

The $U_i$-s thus cover alternative explanations to SUSY and GUT for $E_i$-s. $E_4$ and $E_6$ are fully explained by SM and therefore need no alternative explanations. We therefore need to introduce $U_1, U_2, U_3$ and $U_5$. As we will se below, it makes sense to distinguish two different types of alternative explanation for $E_3$.

Theories covered by $U_i$ must be compatible with $H_{SM}$.  Often, theories covered by a  $U_i$ will also be compatible with SUSY, GUT, or all four basic hypotheses $H_k$ and therefore can be added to them. The $U_i$-s can be combined to cover full hypotheses that can explain several or even all empirical evidence we consider. A fully spelled out hypothesis is characterized by the conjunction of a basic hypothesis $H_k$ and an allowed combination of $U_i$-s.
For $U_i$-s compatible with a basic hypothesis $H_k$, we assume that credence in  $U_i$ is independent of credence in $H_k$:

\begin{equation}
P(H_{k}, U_i) = P(H_{k})P(U_i)
\end{equation}

We  thus have:

\begin{equation}
P(E_i|H_{k}) = (P(E_i|H_{k}, \neg U_i)(1 - P(U_i)) + (P(E_i|H_{k}, U_i)P(U_i)
\end{equation}

\subsection{Likelihoods and Priors}

The updating of a basic hypothesis such as $H_{SUSY}$ under a given set of evidence $E_i$ depends on (i) the likelihoods  $P(E_i|H_k)$ and $P(E_i|U_i)$, and (ii) the  priors $P(H_k)$ and  $P(U_i)$. All of these entries need to be specified to get a quantitative result for the posteriors of $H_{SUSY}$.

In the following, we will assess the likelihoods for the basic hypotheses without $U_i$-s added. We will denote those hypotheses as SM, SUSY, GUT, and SUSYGUT, to distinguish them from the $H_k$-s, which cover added $U_i$-s.

\subsubsection{Specifying the Likelihoods}

\noindent
\textit{Data predicted by hypotheses}
\vspace{3mm}

We have formulated the pieces of empirical evidence $E_i$ in a way that resembles the theoretical predictions of the theories we aim to discuss. Therefore, we set the $P(E_i|H_k)$ close to 1 for the basic theories that explain it. Likelihoods that fall into this category are: 

\vspace{3mm}

$E_1$ is implied by GUT and SUSYGUT.

\vspace{3mm}

$E_2$ is implied by SUSYGUT.

\vspace{3mm}

$E_4$ is strongly favored by SM and SUSY.

\vspace{3mm}

$E_5$ is implied by SM and GUT.

\vspace{3mm}

\noindent
\textit{Data consistent with hypotheses}
\vspace{3mm}

There is data that is not favored by certain hypotheses but is, a priori, a plausible option. Those hypotheses get assigned a mid range likelihood. 
\vspace{3mm}

$E_2$ is not considered a fully expected pattern of coupling constant values based on GUT (that is, GUT without SUSY), but it may be considered a possible pattern. We assume a likelihood

\begin{equation}
P(E_2|H_{GUT})=0.2.
\end{equation}

\vspace{3mm}

\vspace{3mm}

$E_4$ is at variance with some parts of the SUSYGUT parameter space but is consistent with others. We assume an overall likelihood 

\begin{equation}
P(E_4|H_{SUSYGUT})=0.5.
\end{equation}

\vspace{3mm}

$E_5$ is consistent with SM, but is not favored by it over moderately higher mass values. $E_5$ is more probable assuming SM-GUT. This is because a GUT theory that has gauge couplings unify in a perturbative fashion also requires all other couplings in the theory to be perturbative. The Higgs boson self coupling can only stay perturbative up to the GUT scale if the Higgs boson mass is much lighter ($< 170\, {\rm GeV}$) than it otherwise could be and still be a respectably defined particle ($\sim 700\, {\rm GeV}$). We are somewhat conservative in this preference of light Higgs mass ($E_5$) for SM-GUT ($P(H|E_5)=0.6$) over SM without GUT ($P(H|E)=0.3$), given it only a factor of two higher likelihood, whereas some might given the likelihood ratio even higher.

So we have:

\begin{eqnarray}
P(E_5|H_{SM})  = 0.3 \\
 P(E_5|H_{SM-GUT}) = 0.6.
\end{eqnarray}

\vspace{3mm}

$E_6$ stands in tension with a wide range of low energy SUSY models, but is to be expected based on others, such as split SUSY models (see~\cite{Wells:2003tf,Arkani-Hamed:2004ymt,Arkani-Hamed:2004zhs,Wells:2004di}). Therefore, SUSY and SUSYGUT get mid range likelihoods. 
Note that we specify priors here. That is, we specify credences before taking into account other evidence, such as the observation of the separation of scales. That evidence does have a disfavoring effect on split SUSY models compared to non-split SUSY when updated on. But it does not inform the likelihoods that are specified based on priors. We thus assume a priori that split SUSY is equally probable as non-split low energy SUSY, which leads to 

\begin{equation}
P(E_6|H_{SUSY})  = P(E_6|H_{SUSYGUT}) = 0.5
\end{equation}

\vspace{5mm}
\noindent
\textit{Data slightly suppressed by hypotheses}

\vspace{3mm}

$E_1$ is fairly unlikely under SM and SUSY, as discussed in Section 3. We have

\begin{eqnarray}
P(E_1|H_{SM})  = 0.032 \\
 P(E_1|H_{SM-GUT}) = 0.01
\end{eqnarray}

\vspace{3mm}

$E_3$ can, to some extent, be explained by low energy SUSY due to the cancellation of quadratic contributions. There remains a logarithmic suppression of 1/16 of the observed $10^{-16}$ scale difference, however. Taking into account the fact that we assigned a prior of 0.5 to split SUSY, which does not reduce the finetuning needed for the separation of scales to a significant extent, we need to add a factor 1/2 to the credence and end up with 

\begin{equation}
P(E_3|H_{SUSY})=P(E_3|H_{SUSYGUT}) = 1/32
\end{equation}

\vspace{3mm}

$E_4$ rules out a large part, though not all, of non-supersymmetric GUT parameter space. We assume

\begin{equation}
P(E_4|H_{SUSY}) = 0.1
\end{equation}

\vspace{5mm}
\noindent
\textit{Data in strong tension with hypotheses}
\vspace{3mm}

$E_2$ is highly unlikely under SM and SUSY. As discussed in Section 3, we have

\begin{eqnarray}
P(E_2|H_{SM})  =  P(E_2|H_{SUSY}) = 10^{-3}
\end{eqnarray}

\vspace{3mm}

$E_3$ is extremely unlikely under SM and GUT if one treats parameter selection as a random pick from parameter space with a linear measure and gets a minimal likelihood. (Deviating from that understanding will be discussed in the contexts of $U_i$-s.)

\vspace{5mm}
\noindent
\textit{Likelihoods for $U_i$-s}
\vspace{3mm}

Each of the hypotheses $U_1-U_3, U_5$ gets likelihood 1 for the evidence the explanation of which it covers. With respect to other evidence, the $U_i$-s do not alter the likelihoods of the basic hypothesis when added to it. 

Evidence $E_3$ requires a little further attention. SUSY and SUSYGUT do not predict $E_3$ but only reduce a linear to a logarithmic suppression of the observed separation of scales.  $U_3$ therefore covers the only (albeit unconceived) full explanation of $E_3$ that is considered in our scheme. It is plausible, however, to consider a scenario that rejects the linear measure on the parameter space for energy scales without providing a full explanation of the observed separation of scales. This view amounts to rejecting the significance of the naturalness argument rather than having a full explanation of the observed generation of scales. It implies the same likelihood as low energy SUSY and needs to be covered by a different hypothesis on an alternative explanation:

\vspace{3mm}

\noindent
$U_{3N}$: The appropriate probability measure on the parameter space that controls the setting of the electroweak scale is logarithmic. 
\vspace{3mm}

We then have $P(E_3|U_{3N})= 1/16$.

\subsubsection{Constructing the full set of theories}

As described above, we have four ``core theories", which we call SM (Standard Model), SUSY (minimal supersymmetric model), SM-GUT (SM base theory embedded within a grand unified theory), and SUSY-GUT (supersymmetric theory embedded within a grand unified theory). Each of these four theories has  five unconceived alternative additions: $U_1$, $U_2$, $U_{3}$, $U_{3N}$, and $U_{5}$. The meaning of each is
\begin{itemize}
\item[$U_1$:] unconceived explanation for why fermions should necessarily fall within GUT representations.
\item[$U_2$:] unconceived explanation for why gauge couplings should unify well in the UV
\item[$U_3$:] unconceived explanation for the large hierarchy between $M_{\rm Planck}$ and $M_{\rm EW}$
\item[$U_{3N}$:] unconceived explanation that the appropriate probability measure on the parameter space of the electroweak scale is logarithmic, and thus $P(E_3|u_{3N})=1/16$.
\item[$U_5$:] unconceived explanation that guarantees light Higgs boson, $m_H<150\,{\rm GeV}$
\end{itemize}

From these four core theories and the additional five unconceived alternative additions that can be given to each theory, we form a total of 128 theories. The total number of theories results from each of the four core theories having $2^5=32$ permutations of unconceived alternatives. For example, for the SM, one has SM, SM+U1, SM+U2, \ldots, SM+u1+U2, \ldots, SM+U2+U5, \ldots, SM+U1+U2+U3+U3N+U5.

The priors for each of these 128 theories is determined by 7 inputs: 
\begin{itemize}
\item[$x_{\rm SM}$:] Probability of SM being correct and SUSY incorrect
\item[$x_{\rm GUT}$:] Probability of GUT being correct
\item[$\tilde u_1$:] Probability factor for $U_1$ unconceived alternative
\item[$\tilde u_2$:] Probability factor for $U_2$ unconceived alternative
\item[$\tilde u_3$:] Probability factor for $U_3$ unconceived alternative
\item[$\tilde u_{3N}$:] Probability factor for $U_{3N}$ unconceived alternative
\item[$\tilde u_5$:] Probability factor for $U_5$ unconceived alternative
\end{itemize}
What is meant by ``probability factor for $U_i$" is that the core theory gets multiplied by this factor when determining the prior. For example the prior for $SM+U_1+U_5$ is $x_{\rm SM}\tilde u_1\tilde u_5$. As another example, the prior for $\text{SUSY-GUT}+U_1+U_2+U_{3N}$ is $x_{\rm SUSY}x_{\rm GUT}\tilde u_1\tilde u_2\tilde u_{3N}$, where $x_{\rm GUT}=1-x_{\rm SM}$. The priors for all of these theories computed this way are $\tilde P_\alpha$, where $\alpha=1\ldots 128$ are then renormalized:
\begin{equation}
P_\alpha=\frac{\tilde P_\alpha}{\sum \tilde P_\alpha}.
\end{equation}

Table~\ref{table:likelihoods} is a table of ``primitive likelihoods" with respect to the six experimental evidences for the four core theories (the $U_0$ row in each box), along with the additional unconceived alternatives. If we are considering one of the 128 theories that has an unconceived alternative $U_i$ activated that increases the likelihood for $E_i$ then that value of the likelihood is used. For example, if our theory $H$ is $\text{SM}+U_2+U_5$ and we wish to know $P(E_2|H)$, we do not use the standard theory value of $10^{-3}$ but rather $P(E_2|H)=1$, since $U_2$ is an activated unconceived alternative.

\begin{table}[ht]
\small 
\centering
\begin{tabular}{|c|c|c|c|c|c|c|c|}
\hline
$P(E_i|H_\alpha)$  & & $E_1$ & $E_2$ & $E_3$ & $E_4$ & $E_5$ & $E_6$ \\ 
\hline
 SM & $U_0$ &    0.032  & $10^{-3}$   &  $10^{-6}$ & 0.8 & 0.3 & 1  \\ 
  & $U_1$   &  $1$    & $10^{-3}$ &  $10^{-6}$  & 0.8 &  0.3 & 1  \\ 
 & $U_2$ &   0.032      & 1  &  $10^{-6}$  & 0.8 &  0.3 & 1  \\ 
 & $U_{4N}$  & 0.032  & $10^{-3}$  &   0.0625  & 0.8 & 0.3 & 1 \\ 
 & $U_{4C}$ &   0.032  &  $10^{-3}$ &    1  & 0.8 & 0.3 & 1   \\ 
 & $U_{K6}$ & 0.032  & $10^{-3}$ &   $10^{-6}$  & 0.8 & 1 & 1\\ 
\hline
SUSY & $U_0$  & $10^{-2}$  & $10^{-3}$ &  0.0325  & 0.8& 0.9  & 0.5  \\ 
  & $U_1$  &  1 & $10^{-3}$ & 0.0325  & 0.8 & 0.9 & 0.5  \\ 
 & $U_2$ &  $10^{-2}$ & 1    & 0.0325  & 0.8 & 0.9 & 0.5    \\ 
 & $U_{4N}$ & $10^{-2}$ & $10^{-3}$  & 0.0625  & 0.8 & 0.9 & 0.5  \\ 
 & $U_{4C}$ & $10^{-2}$ & $10^{-3}$ & 1  & 0.8 & 0.9 & 0.5  \\ 
 & $U_{K6}$ & $10^{-2}$ & $10^{-3}$  & 0.0325  & 0.8 & 1 & 0.5  \\ 
\hline
SM-GUT & $U_0$  & 1 & 0.2   & $10^{-6}$ & 0.1 & 0.6 & 1  \\ 
  & $U_1$  & 1 &  0.2 & $10^{-6}$ & 0.1  & 0.6 & 1  \\ 
 & $U_2$ &  1 &  1 &  $10^{-6}$ & 0.1  & 0.6 & 1  \\ 
 & $U_{4N}$ & 1  &  0.2 &   0.0625 & 0.1  & 0.6 & 1  \\ 
 & $U_{4C}$ & 1 &  0.2 &  1 & 0.1   & 0.6 & 1  \\ 
 & $U_{K6}$ & 1 &  0.2  &  $10^{-6}$ & 0.1   & 1 & 1  \\ 
\hline
SUSY-GUT & $U_0$  & 1 & 1  & 0.0325 & 0.5 & 0.9 & 0.5 \\ 
  & $U_1$  & 1 & 1 &  0.0325 & 0.5 & 0.9 & 0.5   \\ 
 & $U_2$ & 1 & 1 &  0.0325 & 0.5 & 0.9 & 0.5  \\ 
 & $U_{4N}$ & 1  & 1 & 0.0625 & 0.5 & 0.9 & 0.5    \\ 
 & $U_{4C}$ & 1 & 1 &   1 & 0.5 & 0.9 & 0.5  \\ 
 & $U_{K6}$ & 1 & 1 &  0.0325 & 0.5 & 1 & 0.5  \\ 
\hline
\end{tabular}
\caption{\label{table:likelihoods} Likelihood table.}
\end{table}

From the Priors and the Likelihood table we are able to find posteriors for all 128 theories after applying all the $E_i$ experimental evidences. The sum of the posteriors will automatically sum to 1, as is standard in the Bayesian analysis.

In the text we present some posteriors that reflect all the theories with some property. We obtain such a posterior by summer over all theories with the stated property. For example, we determine the posterior for ``SM-like" theories by summing over the posteriors of all theories that are ``SM-like" including over GUT version of the SM and the various permutations of unconceived alternatives attached to the SM-like theories. SUSY-like theories posterior is the complement of ``SM-like" theories. The posterior for GUTs is the sum over all GUT theories, including SM-like GUTs and SUSY GUTs. ``Pure SM" with no supersymmetry and no GUTs is the sum over the first 32 theories that includes the ``SM" and all the permutations of unconceived alternatives.

\subsection{Specifying the priors}

\subsubsection{Specifying informed priors for low energy SUSY and GUT}

Though priors of scientific hypotheses contain a subjective element, they are not arbitrary. A rough initial guideline for selecting priors is based on general considerations about the research process\footnote{For a specific analysis of this kind of reasoning in terms of meta-empirical assessment, see e.g. \cite{Dawid2018} }.  In light of the general success rate of scientific theories in mature science, a general principle seems conducive to a plausible assessment of the given theory's prospects of being viable: if a theory has not found empirical or contextual support but appears to be consistent with the known empirical data, has no profoundly troublesome characteristics and faces no seemingly lethal conceptual problems, it should not be considered likely but should be taken seriously as a possibility. In our Bayesian analysis, we will translate this attitude into the principle that a theory like low energy SUSY, before taking conceptual and empirical arguments for or against it into account,  should be given a prior credence of one to three percent.\footnote{Note that this is the credence that the theory is viable in some empirical domain rather than the credence in the absolute truth of the hypothesis. The latter could still be set to be very low for philosophical reasons.}

Starting from this prior, each of the arguments 1-4 of Section 2.2. will be assumed to increase credence by some moderate margin, without offering a detailed statistical analysis. 
The esthetic argument and the argument from theoretical embedding (arguments 1-3 of Section 2.2), are at their core arguments for SUSY rather than for low energy SUSY. Nevertheless, each conceptual argument that favors the prospect that SUSY plays a role at high energies would also justify a slightly higher credence in low energy SUSY. It seems fair to say that the three arguments in conjunction do generate some increase of credence in low energy SUSY in the eyes of most high energy physicists. 

Argument 4 is based on an observation that may be framed as slight empirical confirmation of low energy SUSY: Low energy SUSY renders plausible the existence of stable particles that would contribute to deviations from Baryonic matter predictions on galaxy rotation curves and the larger scale cosmological dynamics. A wide  range of collected data $E_{DM}$ on galaxy rotation curves and other cosmological aspects supports the dark matter hypothesis. In Bayesian terms, this leads to $P(E_{DM}|H_S)>P(E_{DM})$, which in turn implies that  the evidence for dark matter increases the credence in low energy SUSY. The situation is complex and difficult to model in detail, however, for at least three reasons. First, SUSY provides a plausible framework for a dark matter candidate, but it does not strictly imply the existence of a stable lightest supersymmetric particle that could play that role. Second, even assuming that there is a stable lightest SUSY particle, it is difficult to assess the amount of dark matter that would be implied by low energy SUSY. Third, there are known alternative theories that offer dark matter candidates as well (and there are also proposals (albeit of limited popularity) that suggest alternatives to dark matter itself). In the present paper, we therefore avoid a specific model of updating credence in SUSY under cosmological data. We treat dark matter as a component that enters the specification of an informed prior in addition to arguments 1-3. 

Let us now spell out the set of prior credences of three different agents. Agents A and B attribute 2.5\% credence to low energy SUSY just based on the hypothesis' coherence:

\begin{equation}
P_{A}(H_{SUSY})= P_{B}(H_{SUSY}) = 0.025   \label{priorA2}
\end{equation} 
\vspace{5mm}

They increase that credence by a few percent due to esthetic arguments 1 and 2 and argument 3 of theoretical embedding and by a few more percent based on dark matter considerations (argument 4).   Arguments 1-4 thus leaves them, on our modeling, with an informed prior credence:

\begin{equation}
P^I_{A}(H_{SUSY}) = P^I_{B}(H_{SUSY}) =0.1   \label{ipriorA1}
\end{equation} 
\vspace{5mm}

A more skeptical agent C, who starts with a prior a little lower than A and B, attributes less significance to any of the arguments 1-4, and ends up with an informed prior of 5\%. 

\begin{equation}
P^I_{C}(H_{SUSY})=0.05   \label{ipriorA2}
\end{equation} 
\vspace{5mm}

It is instructive to spell out the uninformed priors explicitly in terms of the meta-inductive assessment that motivates them. Agents A and B who attribute a 2.5\% credence to low energy SUSY before considering any of arguments 1-7 in its favor (or any arguments against it), believe that one out of 40 theories that look a priori as promising as low energy SUSY would end up being empirically viable.\footnote{Agents A and B are thus fairly moderate, both with regard to their prior credences and with regard to their assessments of the significance of arguments 1-4 in favor of SUSY. A more outspoken optimist might well endorse higher credences in those respects. However, since we are mostly interested in the extent to which the three arguments for which we spell out a Bayesian updating model can boost credences, we take it to be more instructive to start with moderate or low priors for all modeled agents.} Agent C is more pessimistic that this and assumes that just a little over 1\% of the theories that look a priori as plausible as low energy SUSY would end up being empirically successful. We suggest that going significantly below this pessimistic expectation would have little plausibility. 

Regarding GUT, we will attribute the same uninformed priors as in the case of SUSY. For agents A and B, we thus have: 

\begin{equation}
P_{A}(H_{GUT}) = P_{B}(H_{GUT}) =0.025   \label{priorA2G}
\end{equation} 
\vspace{5mm}

 There is a certain conceptual appeal in support of GUT, roughly comparable to the esthetical arguments in favor of SUSY which justifies a higher informed prior. GUT is not strictly implied by string theory, however, and does not provide a dark matter candidate. We therefore assume informed priors that are lower than those for low energy SUSY:

\begin{eqnarray}
P^I_{A}(H_{GUT}) = P^I_{B}(H_{GUT}) &=& 0.05   \label{ipriorA1G}\\
P^I_{C}(H_{GUT}) &=& 0.025   \label{ipriorA2G}
\end{eqnarray} 
\vspace{5mm}

We will use eqns (\ref{ipriorA1})-(\ref{ipriorA2G}) as informed priors based on which we will model updating under evidence 5-7 as described in the upcoming sections.

 Keep in mind, once again, that the point of our analysis is not to propose or advertise specific values of credence but to develop a better understanding of the way data in conjunction with some background assumptions influence credences and can, under plausible assumptions, lead to fairly substantial posteriors.

\subsubsection{Specifying priors for $U_i$-s}

While $H_k$-s are known hypotheses, the $U_i$-s cover both known and unconceived alternatives. For several of the $E_i$-s, no alternative explanations are known, which means that the corresponding $U_i$ covers only unconceived alternatives.

$U_1$ and $U_2$ point at unconceived explanations for evidences $E_1$ that  particles fit in GUT representations and $E_2$ that the three couplings converge to the extent they do, respectively. No explanation other than GUT has been proposed. As it stands, it seems difficult to imagine how an alternative explanation could work. This does not rule out the existence of an unconceived explanation, but it seems natural, in light of past experience with similar contexts, to attribute a modest probability to this scenario. Our default estimate (assigned to agent $B$) will be to  assign to such $U_i$-s the same prior as to the hypothesis that a unsupported new hypothesis was viable (last section). In a straightforward case such as  $E_1$, where we have a known theory that does fully explain the evidence (GUT), our selection of prior means the following: we  assign the same credence to the existence of an unconceived explanation for $E_i$ that seems difficult to imagine as we would assign to the known explanation (SUSY) if we updated that credence from the uninformed prior just on the given evidence $E_i$. For the pessimistic agent $C$, we double that prior. For the optimistic agent $A$, we half it.\footnote{Note that agents $A$ and $B$ assume the same prior for SUSY and the same prior for GUT. They only differ with respect to their priors for the $U_i$-s, which express  their assessments of the probabilities of unconceived alternatives. As we will see, this difference has quite significant implications for posterior credences.} Based on  eqn (\ref{priorA2G}),  we have:

\begin{eqnarray}
P_{C}(U_{1})= P_{C}(U_{2})&=& 0.05   \label{priorA1U}\\
P_{B}(U_{1})= P_{B}(U_{2})&=& 0.025   \label{priorA2U}\\
P_{A}(U_{1})= P_{A}(U_{2}) &=& 0.0125   \label{priorA3U}
\end{eqnarray} 
\vspace{5mm}

In the case of separation of scales ($E_3$), anthropic reasoning might in principle offer a framework for explaining the phenomenon. The open questions that arise in that context reduces credence in the hypothesis that a full explanation of separations of scales exists in that framework. While one might still attribute a higher prior to $U_3$ than to $U_1$ and $U_2$ on that basis, we will, for the sake of simplicity, assume the same priors as before.

\begin{eqnarray}
 P_{C}(U_{3})&=& 0.05   \label{priorA1U}\\
 P_{B}(U_{3})&=& 0.025   \label{priorA2U}\\
 P_{A}(U_{3})&=& 0.0125   \label{priorA3U}
\end{eqnarray} 
\vspace{5mm}

$P(H_{3N})$ does not point at an alternative hypothesis but at the view that one should not apply a linear probability measure to the parameter of energy scales. If we apply a logarithmic measure instead, naturalness stops being an argument for low energy SUSY. Since the idea to refrain from applying a linear measure is a matter general attitude towards current physics rather than of imagining an unconceived alternative to known theories, we must not assign it the low prior credence that seemed adequate in cases $U_1-U_3$. Rather,we assume that the pessimist C sees no basis for accepting or rejecting the linear measure and therefore assigns credence 0.5 to each of them.  Agent B takes the naturalness argument to be quite significant though not uncontested and the optimist agent A takes it to be fairly reliable. These views are represented by correspondingly lower credences in the hypothesis that one should apply a logarithmic rather than a linear probability measure.

In addition, we model an agent $B^{*}$ who shares all credences except for those involving $U_3$ and $U_{3N}$ with agent $B$. But agent $B^{*}$  has doubts whether finetuning arguments are good guidelines towards new physics ($\rightarrow$ higher $P(U_{3N})$) and, moreover, attributes a considerable credence to the idea that, if there is an explanation of the separation of scales at  all, an anthropic argument could provide that explanation ($\rightarrow$ higher $P(U_3)$). We thus model agent $B^{*}$ to share the prior credences  $P(U_3)$ and $P(U_{3N})$ with agent $C$. 
We set:

\begin{eqnarray}
 P_{C}(U_{3N})= P_{B^{*}}(U_{3N}) &=& 0.5   \label{priorA1U}\\
 P_{B}(U_{3N})&=& 0.25   \label{priorA2U}\\
 P_{A}(U_{3N})&=& 0.125   \label{priorA3U}
\end{eqnarray}

\vspace{3mm}

$U_5$ differs from $U_1-U_3$ since there are known alternative hypotheses that put stronger constraints on the Higgs mass than the standard model. Credence in those known alternatives hypothesis, once again, depends on the generic willingness to have trust in a hypothesis, plus on  specific arguments that speak in favor or against those theories. Since we take it that none of those theories has empirical support comparable to SUSYGUT, and since it makes no sense to enter into the details of each of those models, we assume a credence in the known models that is lowest for the pessimist C and highest for the optimist A. On the other hand, there may also be unconceived alternatives that constrain the Higgs mass. The pessimist has higher credence in such unconceived alternatives than the optimist. In conjunction it therefore seems plausible to assume a universal credence in $U_5$ to all agents. We will assume:

\begin{eqnarray}
 P_{C}(U_{5})=
 P_{B}(U_{5})=   
 P_{A}(U_{5})= 0.1   \label{priorA3U}
\end{eqnarray} 
\vspace{3mm}

\section{Running the Model}

We are now in the position to run the model for the four agents, $A$, $B$, $B^{*}$ and $C$. 
Results are invariant under permutation of the individual steps of updating under $E_1-E_6$. 
We are particularly interested in two posteriors: credence before LEP2 and the LHC and credence that accounts for LHC data. 

\vspace{3mm}

{\bf Credence before the LHC}:  Updating under $E_1-E_4$ gives us roughly a posterior before the runs of the LEP2 and the LHC.\footnote{This construal seems instructive despite some slight chronological inaccuracies involved.}
We get the following posteriors: 

\begin{eqnarray}
 P_{C}(H_{GUT}|E_1,..E_4)=0.32 \\
 P_{B^{*}}(H_{GUT}|E_1,..E_4)=0.80\\
 P_{B}(H_{GUT}|E_1,..E_4)=0.83 \\
 P_{A}(H_{GUT}|E_1,..E_4)=0.94
\end{eqnarray} 

\begin{eqnarray}
 P_{C}(H_{SUSY}|E_1,..E_4)=0.22 \\
 P_{B^{*}}(H_{SUSY}|E_1,..E_4)=0.63 \\
 P_{B}(H_{SUSY}|E_1,..E_4)=0.69 \\
 P_{A}(H_{SUSY}|E_1,..E_4)=0.82
\end{eqnarray} 

\begin{eqnarray}
 P_{C}(H_{SUSYGUT}|E_1,..E_4)=0.19 \\
 P_{B^{*}}(H_{SUSYGUT}|E_1,..E_4)=0.62 \\
 P_{B}(H_{SUSYGUT}|E_1,..E_4)=0.67 \\
 P_{A}(H_{SUSYGUT}|E_1,..E_4)=0.82
\end{eqnarray} 
\vspace{5mm}

{\bf Credence after the LHC}: Updating under $E_1, E_2, E_4, E_5, E_6$ roughly gives us the posterior in light of the LHC data. $E_4$, the separation of scales, has stopped being a factor of discrimination between SUSY and SM since split SUSY, which survives the LHC data, does not provide a solution to the naturalness problem. Though it is true that the fine-tuning is smaller for split SUSY than for SM, it is still substantially higher than the 1-5\% probability of there being an unconceived explanation of the separation of scales. The effect the naturalness argument has on the credence in split SUSY thus is negligible. Since none of the known theories provide an explanation of the separation of scales, differences in views on the naturalness argument become irrelevant and $B^{*}$ has the same set of credences as $B$.

 \vspace{3mm}

\begin{eqnarray}
 P_{C}(H_{GUT}|E_1, E_2, E_4,..E_6)=0.38 \\
 P_{B}(H_{GUT}|E_1, E_2, E_4,..E_6)=0.82 \\
 P_{A}(H_{GUT}|E_1, E_2, E_4,..E_6)=0.92
\end{eqnarray} 

\begin{eqnarray}
 P_{C}(H_{SUSY}|E_1, E_2, E_4,..E_6)=0.19 \\
 P_{B}(H_{SUSY}|E_1, E_2, E_4,..E_6)=0.54 \\
 P_{A}(H_{SUSY}|E_1, E_2, E_4,..E_6)=0.61
\end{eqnarray} 

\begin{eqnarray}
 P_{C}(H_{SUSYGUT}|E_1,..E_4)=0.17 \\
 P_{B}(H_{SUSYGUT}|E_1,..E_4)=0.53 \\
 P_{A}(H_{SUSYGUT}|E_1,..E_4)=0.60
\end{eqnarray} 
\vspace{5mm}

If we compare credences before and after the LHC, we thus find the following trajectories:

\begin{eqnarray}
 P_{C}(H_{GUT}): 0.32 \rightarrow 0.38 \nonumber\\
 P_{B^{*}}(H_{GUT}): 0.80 \rightarrow 0.82 \nonumber\\
 P_{B}(H_{GUT}): 0.83 \rightarrow 0.82 \nonumber\\
 P_{A}(H_{GUT}): 0.94 \rightarrow 0.92 \nonumber
\end{eqnarray} 

\begin{eqnarray}
 P_{C}(H_{SUSY}): 0.22 \rightarrow 0.19 \nonumber \\
 P_{B^{*}}(H_{SUSY}): 0.63 \rightarrow 0.54 \nonumber \\
 P_{B}(H_{SUSY}): 0.69 \rightarrow 0.54 \nonumber\\
 P_{A}(H_{SUSY}): 0.82 \rightarrow 0.61 \nonumber
\end{eqnarray} 
\vspace{5mm}

\section{Conclusions}

The Bayesian analysis leads to the following general conclusions. We identified general principles of pessimistic or optimistic attitudes in the assessment of theories in the absence of direct empirical testing of the theory's core predictions. Consistently adhering to those attitudes puts some degree of constraint on the choice of priors. The scientific record in a research field renders overly optimistic or pessimistic attitudes implausible, which further constrains the spectrum of reasonable priors. Updating within a plausible framework of priors and assessments of the probabilities of unconceived alternatives allows one to identify a range of  credences one may take to be plausible. 

As expected, we still get a wide spectrum of posteriors depending on whether one adopts a more optimistic or more pessimistic attitude towards theory building. We aimed at selecting priors that seemed to us within the range of what seems reasonable and justifiable in light of previous experiences of success and failure in high energy physics. We would suggest that priors significantly below what we assigned to the pessimist would be difficult to defend in that light. On the other hand, we would argue that our optimistic agent A, remains well within the range of plausible prior assumptions.

On our analysis, the main driver of credence in SUSY is the strong support of GUT. GUT representations ($E_1$) and the meeting of the gauge couplings $(E_2)$ generate high credence in GUT. The fact that the meeting of the gauge couplings works much better in a SUSY framework generates substantial credence in SUSY via GUT. 
The quantitative analysis shows that fairly high credence in GUT is achievable already based on conservative prior assumptions. It can reach beyond the 90\% range based on optimistic though justifiable priors.\footnote{Interestingly, the pessimist's credence in GUT, though still moderate, has increased in light of LHC data, due to the fact that low Higgs masses are more probable in SM-GUTs than in the SM.} Credence in SUSY is lower but can become quite substantial as well. A pessimistic though justifiable account leads to credences that remain significant but do not amount to substantial commitment to theories' viability.  

What may be the most significant result of this analysis is concerned with the change in credences from a pre-LHC to a post-LHC perspective. The variation with respect to these changes is very significantly constrained if one requires that agents consistently apply their optimistic or pessimistic attitudes. To give an example, if they take very seriously the reduction of parameter space for low energy SUSY, they should not, without good reason, ignore the reduction of the range of allowed Higgs mass values by SUSY. If one applies the principle that attitudes should be consistent across the appraisals of various sets of data,  one does find a reduction in credence based on the LHC data. However, that reduction is fairly moderate and remains well within the range of what seemed to be a justifiable spectrum of credences before the LHC. A serious reduction in credence in SUSY is experienced by the optimist, whose doubts about SUSY have doubled due to the collapse of the naturalness argument and the failure to find SUSY signatures. Agent B experiences a significant but moderate reduction in credence. The pessimist and the agent B-type skeptic about naturalness-arguments experience only next to negligible reductions in credence. For them, the predictive significance of the correct Higgs mass nearly compensates for the loss of low energy SUSY parameter space without split SUSY structure. 

The Bayesian analysis points at a couple of potential sources of intuitive overestimation of the difference between plausible pre-LHC and post-LHC credence in low energy-SUSY. Naturalness arguments may be considered very powerful by observers due to the extremely high numbers that characterise the observed finetuning. The Bayesian analysis makes explicit, however, that the significance of an argument of extreme finetuning is controlled entirely by the threat of unconceived alternatives rather than by the degree of extreme finetuning. In a nutshell, given that it is seldom adviseable in a progressing research context to attribute a probability below 1\% to the possibility that an unknown explanation of a given phenomenon emerges, it is of little relevance whether the finetuning to be explained is of the order $10^4$ or of the order $10^{16}$. 

The significance of a naturalness argument for SUSY is further reduced by the fact that SUSY does not fully explain the separation of scales while a possible unconceived explanation (or a vaguely perceived anthropic explanation) might do so. The confirmation value of the measured Higgs mass, on the other hand, may sometimes be underestimated. Since its general significance is reduced in part due to known alternative explanations (credence in which is considered low by the pessimist), the argument does not lose significance if viewed from a pessimist perspective, unlike other arguments in favor of SUSY. 

Finally, the probabilistic analysis suggests that GUT provides the strongest argument in favor of low energy SUSY pre- as well as post-LHC. The GUT-based argument may sometimes be underestimated due to its conditional character. After all, its support is based on a theory that has not found empirical confirmation itself. The high credence in GUT that is generated from fairly low priors in view of empirical evidence nevertheless turns GUT into a strong argument for SUSY.

\section{Acknowledgements}
This research was in part funded by the Swedish Research Council grant number  2022-01893\_VR and by the Wenner-Gren grant GFOh2019-0015.

\bibliographystyle{abbrvnat}
\bibliography{SUSYandBayes}
\end{document}